\documentclass{article}
\usepackage{arxiv}
\usepackage{graphicx}
\usepackage{doi}
\usepackage{cite}

\title{Security and Privacy-Preservation of IoT Data in Cloud-Fog Computing Environment}

\author{ \href{https://orcid.org/0000-0002-9319-3408}{\includegraphics[scale=0.06]{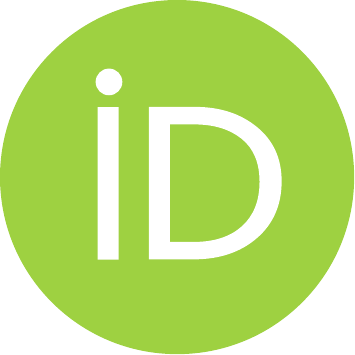}\hspace{1mm}Jatinder Kumar}\thanks{The authors would like to thank National Institute of Technology Kurukshetra, India for financially supporting this research work.} \\
	Department of Computer Applications\\
    National Institute of Technology\\
	Kurukshetra, India \\
	\texttt{jatinder\_61900097@nitkkr.ac.in} \\
	\And
	\href{https://orcid.org/0000-0002-8053-5050}{\includegraphics[scale=0.06]{orcid.pdf}\hspace{1mm}Ashutosh Kumar Singh} \\
		Department of Computer Applications\\
      National Institute of Technology\\
	Kurukshetra, India \\
	\texttt{ashutosh@nitkkr.ac.in} }

\renewcommand{\shorttitle}{\textit{Security and Privacy-Preservation of IoT Data in Cloud-Fog Computing Environment}}

\hypersetup{
pdfsubject={q-bio.NC, q-bio.QM},
pdfauthor={David S.~Hippocampus, Elias D.~Striatum},
pdfkeywords={First keyword, Second keyword, More},
}

\begin{document}

\maketitle

\begin{abstract}
IoT is the fastest-growing technology with a wide range of applications in various domains. IoT devices generate data from a real-world environment every second and transfer it to the cloud due to the less storage at the edge site. An outsourced cloud is a solution for handling the storage problem. Users' privacy can be exposed by storing the data on the cloud. Therefore, we propose a Private Data Storage model that stores IoT data on the outsourced cloud with privacy preservation. Fog nodes are used at the edge side for data partition and encryption. Partitioned and encrypted data is aggregated with the help of homomorphic encryption on the outsourced cloud. For secure query processing and accessing the data from the outsourced cloud, the introduced model can be used on the outsourced cloud. 
\end{abstract}
%https://www.overleaf.com/project/6129194805bd1181f52dde18

% keywords can be removed
\section{Introduction}
With the advancement and deployment of the internet all over the globe, the Internet of Things (IoT) revolutionized our daily lifestyle by providing flexibility and convenience \cite{swain2022efficient,kumar2020ensemble,singh2022privacy11,saxena2022fault,gupta2022auxiliary}. IoT connects billions of machines with the internet and is capable of collecting $\&$ sharing real-time data \cite{chhabra2020secure,saxena2021energy,gupta2020framework}. Major applications of IoT devices include smart homes \cite{khan2016internet,chhabra2020security}, smart cities \cite{pouryazdan2016anchor}, smart healthcare \cite{tripathi2020review,lin2009sage,tripathihedcm}, and smart grid \cite{lu2016privacy,singh2019stock,gupta2019layer,agarwal2019authenticating}. Sensors of IoT devices collect data from the environment and make them digital Intelligence devices \cite{gupta2020seli,kumar2021discussion,singh2020online,gupta2019dynamic,godha2019architecture,nader2015designing,kumar2020decomposition}. For example, a smart vehicle’s sensors sense the environment for traffic for every millisecond \cite{kumar2021performance1,chhabra2021dynamic,saxena2020communication}. By analyzing this information, the vehicle detects the empty road and then decides to speed up or down \cite{hura2020advances,gupta2020guim}. IoT makes the world more responsive and smarter by merging the digital and physical worlds \cite{gupta2020integrated,tiwari2021credit,mittal2021study,kumar2020adaptive}.

It is no doubt that we have great benefits from IoT devices \cite{taneja2015preserving,gupta2022hisa}. However, there are some issues with their benefits, like low computation power and storage problems of IoT devices \cite{chhabra2019dynamic,gupta2022pca,yadav2019advancements}. Local machines are not able to store all data which IoT generates \cite{saxena2021secure,gupta2022tidf,singh2021quantum}. This problem may be solved by cloud computing after outsourcing the data. But the cloud is not secure from the point of privacy. Many attackers may access the data from the clouds \cite{saxena2022intelligent,kaur2017comparative}. One of the famous attacks was the Apple iCloud breach \cite{lewis2014icloud}, where private photos of 500 celebrities leaked. Due to these attacks, the outsourced clouds are always not trustworthy \cite{kumar2019cloud,martini2013cloud,gupta2019confidentiality,deepika2020review,chauhan2020survey,chhabra2019optimal}. As the IoT devices are in a large number indicates that more security and protection are needed \cite{sicari2015security,kolias2017ddos,yang2017survey}. The whole data is stored on the cloud, and when the attackers breach the security, they access the complete data \cite{singh2018web,gupta2022mlrm,sharma2019fast,leng2012link,tagliacane2016network,goh2015comprehensive}. Tian Wang et al. \cite{wang2018three} proposed a scheme that stores the data and in which whole data is not stored only on the cloud.
Data is partitioned and stored at different levels so that the attacker can not access the complete data. IoT devices can not partition data before storing on the outsourced cloud because of less computation power. We have introduced the fog nodes between cloud and IoT devices to accomplish this task. The only factor differentiating cloud and fog computing is the availability of fog nodes close to the edge device. The concept of fog computing is introduced by Cisco \cite{stojmenovic2014fog} and can be abbreviated as “From cOre to edGe” \cite{gupta2022privacy1,sarkar2015assessment,gupta2022privacy,singh2019sql}. As in the middle of edge devices and cloud nodes, fog nodes can compute and transmit data to different levels. Providing fog computing near IoT devices improves latency and real-time computational capabilities of data generated by these devices. The Fog node’s computational capabilities can be used for data partitioning and encryption before sending them into the outsourced cloud \cite{kumar2020cloud}. After storing, data on the outsourced cloud in encrypted form makes statistics a problematic task. Some recent works proposed their schemes to analyze the data on the outsourced cloud \cite{singh2022metaheuristic,raj2016low,tep2015taxonomy,quick2013cloud,kaur2019digital}, but they have not solved these issues technically.

\begin{figure}[!htbp]
    \centering
    \includegraphics[width=0.95\textwidth]{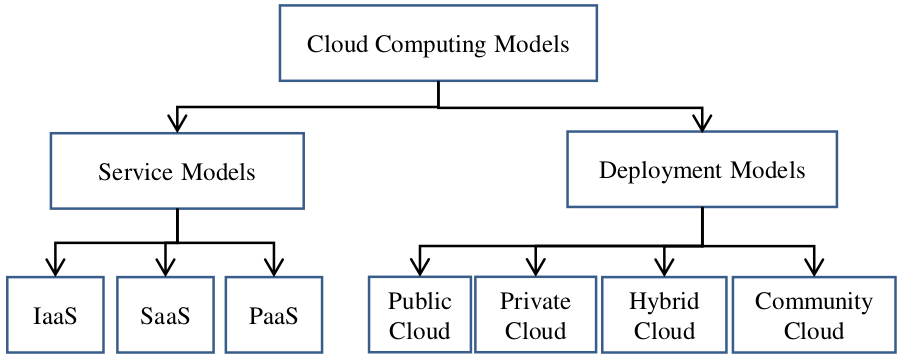}
    \caption{Cloud computing classification}
    \label{fig:cloud}
\end{figure}

The cloud computing is defined as to provide the computing resources over the internet \cite{ge2016evaluating,gupta2020mlpam,gupta2021data,SainiPrivacy,varshney2021machine,yang2018emotion}. These resources can be hardware or software in accordance to their requirements and the architecture of cloud computing is shown in Figure \ref{fig:cloud} on the basis of services and deployment \cite{saxena2021workload,AtmapoojyaPrivacy,saxena2022vm,gupta2022holistic}. The cloud computing can be divided into three main categories according to their services \cite{yadav2022survey,saxena2021survey,makkar2022secureiiot}. These are: Infrastructure-as-a-service (IaaS), Platform-as-a-service (PaaS), and Software-as-a-service (SaaS). On the basis of deployment, the cloud computing is broadly classified into four main categories, i.e., public, private, hybrid, and community cloud \cite{patel2021review,gupta2022compendium,kaur2017data, patel2021lstm,chhabra2018oph,singh2018data,chhabra2018probabilistic,saxena2022high,gupta2018probabilistic,varshney2022plant}.

The benefits of cloud computing include availability, scalability, cost-efficient, mobility, disaster recovery, security, and so on \cite{saxena2018abstract,gupta2022differential,gupta2022quantum,saxena2015ewsa,gupta2022differential1,saxenaa2020communication,kumar2021performance, saxena2021osc,saxena2015highly,saxena2020auto}. These benefits are provided on a pay-as-you-go basis \cite{gupta2022differential4,sharma2021lightweight}. You have to pay only for those services which you are using \cite{saxena2016dynamic,saxena2021proactive,singh2021cryptography,saxena2021op,kumar2020biphase, chhabra2016dynamic,kumar2016dynamic,Goel2022}. With the benefit, there are some challenges also of cloud computing \cite{jalwa2021comprehensive,tiwari2021hybrid,pradhan2021comparative, choudhary2021review,godha2021flooding}. These challenges are data privacy, data security, load balancing, power utilization, resource utilization, and resource migration \cite{saxena2020security,kumar2021self,varshney2021machine1,acharya2021host, singh2022privacy,kesharwani2021real, gupta2022differential2,kumar2021resource}.

The smart grid uses the storage and real-time analysis services of cloud computing. As the cloud is considered semi-trusted, i.e., honest and curious \cite{raj2014low,kumar2010web,gopal2014design,devkota2018image}. The encryption algorithms are used to protect the privacy of customers at the third party (outsourced cloud) \cite{singh2009comparative}. Homomorphic encryption is used to secure and privacy-preserving IoT device data aggregation. The public key of IoT devices is used to encrypt the private data, whereas the private key is used to convert the encrypted data to original data.

In this work, our objective is to achieve the privacy-preservation of data generated by IoT devices and secure query processing on the outsourced cloud. We propose privacy-preserving private data storage model for storing data of IoT devices on outsourced clouds. We use fog nodes for the computation and transmission of data at the edge side. Two outsourced clouds are used to maintain privacy and storage problems. Homomorphic encryption is used to aggregate the encrypted data at clouds. Then a secure device data query scheme is proposed for query processing at the cloud on encrypted data.

\section{Related Work} 

 Andriopoulou et al. \cite{andriopoulou2017integrating} proposed a fog-IoT and cloud-integrated architecture that provides computing services at the end devices side. Healthcare IoT devices collect the sensor data and aggregate it at fog nodes. Fog computing improves latency for the transmission of data to the cloud servers. Therefore, architecture is most suitable for healthcare services. Usman et al. \cite{usman2017sit} proposed a secure IoT architecture having a lightweight encryption algorithm to encrypt IoT data. It uses a 64-bit long key and generates a 64-bit block cipher after 5 rounds of encryption to provide security. An Attribute-Based Encryption (ABE) scheme over data of IoT devices is presented in \cite{wang2014performance}. Key-Policy Attribute-Based Encryption (KP-ABE) and Ciphertext-Policy Attribute-Based Encryption (CP-ABE) schemes are evaluated on different mobile IoT devices that achieve data privacy. Bhandari and Kirubanand \cite{bhandari2019enhanced} presented an advanced cryptographic solution with both asymmetric \& symmetric encryption. The keys are generated using elliptical curve cryptography. This model makes IoT data transmission secure to the cloud. Genitary \cite{gentry2009fully} introduced homomorphic encryption (HE) scheme that allows an arbitrary number of addition and multiplication operations on ciphertexts. However, the computational complexity of these algorithms is too high. A Partial HE scheme that supports ECC is used by \cite{daisy2018adaptable}\cite{vedaraj2021herde} on the integrated framework of IoT and cloud. ECC HE uses a small key size and improves security and ambiguity. Ramesh and Govindarasu \cite{ramesh2020efficient} presented a framework called proxy reciphering as a service and uses FHE \& chameleon hash functions for privacy-preserving even after device keys are compromised. In this work, data is encrypted first with the AES algorithm and then transformed into homomorphic ciphertexts. For the data integration at the cloud, only an addition operation could be sufficient. Paillier cryptosystem is proposed by Pascal Paillier \cite{paillier1999public} that supports an arbitrary number of addition operations and has less computation complexity than FHE.

  The authors proposed secure query processing in WSN to access the user's data on the cloud \cite{shi2010spatiotemporal,miao2020hardness,shi2009secure,singh2022privacy_j}. Yuan et al. present an architecture that enables encrypted query processing over IoT data \cite{yuan2020scalable}. HomomorphicEncryption with random diagonal elliptical curve cryptography integrated with Multi-nomial smoothing Naive Bayes model is proposed by \cite{vedaraj2021herde} over the IoT health cloud system. MSNB model is used for the prediction of diseases with less processing time and low computation cost. In our proposed work, we use Paillier HE for adding the ciphertexts at the outsourced cloud in the proposed PDS model. Our proposed work consists of both the features privacy-preserving model as well as query processing over encrypted data.

\section{Proposed Model}
\begin{figure}[!htbp]
    \centering
    \includegraphics[width=0.75\textwidth]{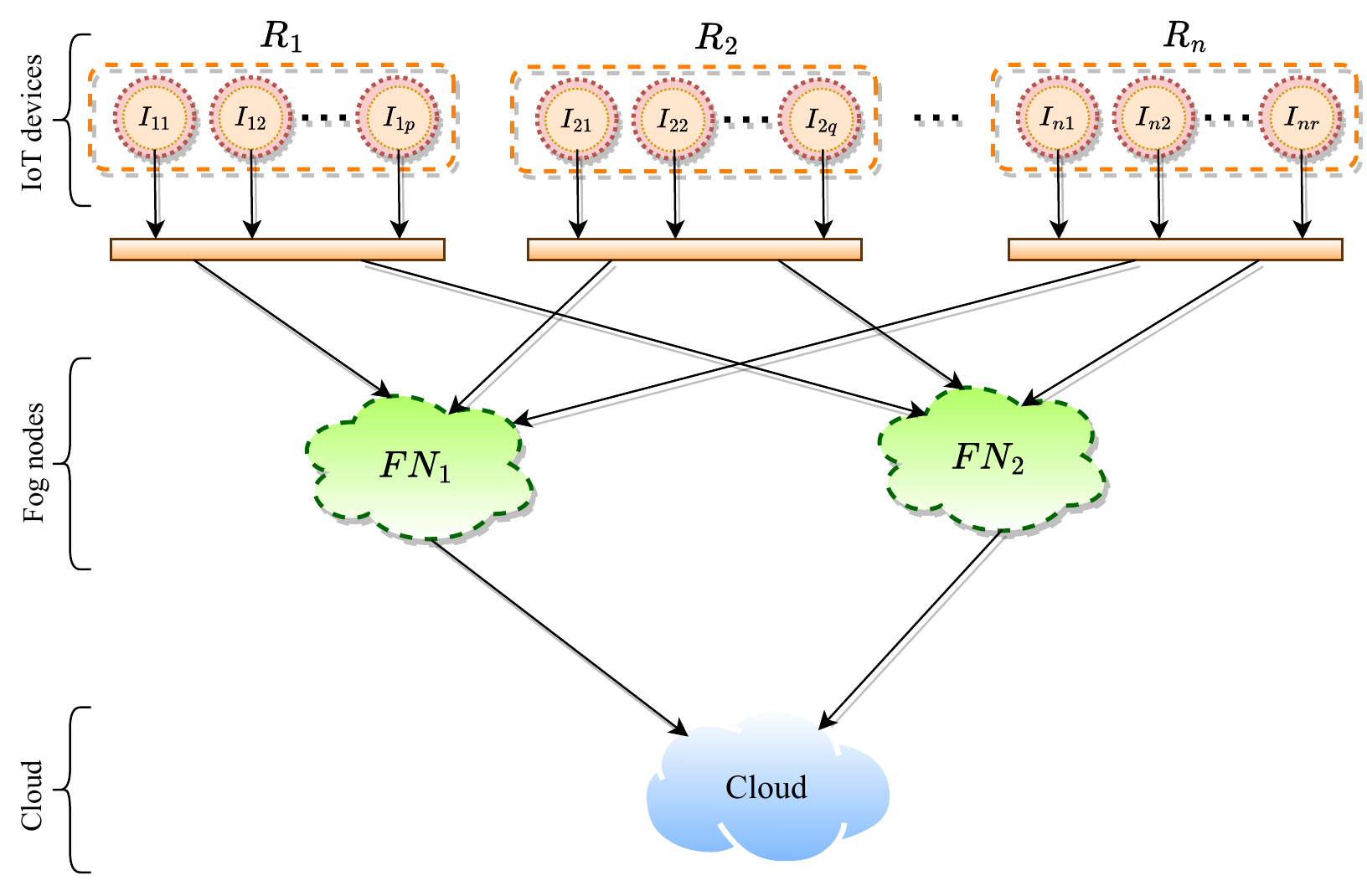}
    \caption{Proposed Model}
    \label{fig:prop}
\end{figure}

The architecture of the proposed model is depicted in Figure \ref{fig:prop}. The proposed model consists three main entities, i.e., IoT devices, fog nodes, and cloud. IoT devices are distributed into $n$ Regions ($R$) and each regions contains IoT devices, such as $R_1$ = \{$I_{11}$, $I_{12}$, ..., $I_{1p}$\}, $R_2$ = \{$I_{21}$, $I_{22}$, ..., $I_{2q}$\}, and $R_n$ = \{$I_{n1}$, $I_{n2}$, ..., $I_{nr}$\}. 

These devices generate data after some period of time ($t$), for example, $I_{11}$ generated data ($D_{11}^t$) at time ($t$). The data report of these devices ($D^t_{Ri}$) = \{$D_{i1}^t$, $D_{i2}^t$, ..., $D_{ix}^t$\} is encrypted with the paillier homomorphic encryption and the encrypted report ($Enc(D^t_{Ri})$) = \{$Enc(D_{i1}^t)$, $Enc(D_{i2}^t)$, ..., $Enc(D_{ix}^t)$\} of IoT devices are forwarded to the corresponding fog node ($FN$). The fog node aggregate  the IoT devices data at different time intervals ($t$ and $\Delta t$) with the property of homomorphic encryption and computes $Enc(D^{t+\Delta t}_{Ri})$ = $Enc(D^t_{Ri})$ + $Enc(D^{\delta t}_{Ri})$. Thenafter, $Enc(D^{t+\Delta t}_{Ri})$ is transferred to the  cloud for storage and analysis. In addition to this, the proposed model can be used for query processing. If data of any IoT device is required, the request of data of the particular device is made to the cloud by providing the identity number. The cloud uses the identity number of devices to fetch the required data in the storage and send the data to the requested authority. The responded data is in ciphertext with the public key of the device owner. So, private key of the corresponding device is required to get the required result. By applying the private key, the requested authority get the required output.

\section{Conclusion}
This paper focuses on the privacy-preserving data aggregation of IoT devices. The IoT devices collect real-time data at different time intervals from the environment. The collected data is encrypted with the Paillier homomorphic encryption at different times. The encrypted data is then after forwarded to the fog nodes. The encrypted data is aggregated at various times and transferred to the outsourced cloud. The proposed model can also be used for secure query processing. Future work can be included the authentication of data at the fog node with the digital signatures.

\bibliographystyle{unsrt}
\bibliography{reference} 
\end{document}